\documentstyle[12pt]{article}

\begin{document}

\centerline{\bf UNIFICATION OF SPINS AND CHARGES IN GRASSMANN
SPACE}

\vspace{3mm}

\centerline{\bf ENABLES UNIFICATION OF ALL INTERACTIONS
\footnote{Talk presented at ICSMP 95, Dubna, July 1995.}}
\baselineskip=16pt
\vspace{0.8cm}
\centerline{\rm NORMA MANKO\v C BOR\v STNIK}
\baselineskip=13pt
\centerline{\it Department of Physics, University of
Ljubljana, Jadranska 19 }
\vspace{0.1cm}
\centerline{\rm and}
\vspace{0.1cm}
\centerline{\it J. Stefan Institute, Jamova 39}
\baselineskip=12pt
\centerline{\it  Ljubljana , 61 111, Slovenia }
\vspace{0.9cm}

\abstract{
In a space of $d $ Grassmann coordinates
two types of generators of  Lorentz transformations can be
defined, one
of spinorial and the other of vectorial character.
Both kinds of operators appear as linear operators in Grassmann
space, definig the fundamental and the adjoint
representations of the group $ SO(1,d$$-$$1) $, respectively. The
eigenvalues of commuting operators belonging to the subgroup
$(SO(1,4)) $ can be identified with spins of either fermionic or
bosonic fields, while the operators belonging to subgroups of $
SO(d$$-$$5) $ ${\supset SU(3)}$  $ { \times SU(2)}$ $ { \times U(1)} $,
determine the Yang-Mills charges.  The theory offers unification
of all the internal degrees of freedom of fermionic and bosonic
fields - spins and all Yang-Mills charges. When accordingly all
interactions are unified, Yang-Mills fields appear as part of the
gravitational field. The theory suggests that elementary
particles are either in the fundamental representations with
respect to the groups determining the spin and  the charges,
or they are in the adjoint representations with respect to the
groups, which determine the spin and the charges.


\vskip 1cm

\noindent
{\bf INTRODUCTION}

\vskip 0.5cm

What today is accepted as elementary particles and fields are either
{\it fermionic fields with the internal degrees of freedom -
spins and all charges} - {\it in the
fundamental representations with respect to the groups} $ SO(1,3),
U(1), SU(2), SU(3) $ or {\it bosonic fields with the internal degrees
of freedom - spins and charges} - {\it in
the adjoint (regular) representations with respect to the same
groups}. Fermions are the Lorentz spinors, forming the
isospin doublets or singlets, the colour triplets
or singlets and the charge singlets. Bosons are the Lorentz
scalars or vectors,
forming the isospin singlets or triplets, the colour
singlets or octets and the charge singlets. There exist no
known fermions with charges in the adjoint
representations and no known bosons with charges in the fundamental
representations. The Higgs scalar appears in the Standard model
as an isospin doublet, but his existence is not yet prooved.

In this talk I am proposing the approach which unifies spins and
charges requiring that (elementary) fermions are in the fundamental
representations with respect to all gauge groups and
(elementary) bosons are in the adjoint representations with
respect to all gauge groups.  The space has $d$ commuting and
$d$ anticommuting ( Grassmann ) coordinates.
All the internal degrees of freedom, spins and charges, are
described by the generators of the Lorentz transformations in
Grassmann space. All gauge fields - gravitational as well as
Yang-Mills including electrodynamics - are defined by
(super)vielbeins.

In Grassmann space there are two types of  generators of
Lorentz transformations and  translations: one is of spinorial
character and determines properties of fermions, the other is of
vectorial character and determines properties of bosons. Both types
of generators are {\it linear differential operators in Grassmann
space}. Their representations can be expressed as monomials of
Grassmann
coordinates $\theta^a$. If $d\geq 15$ the generators of the
subgroup $ SO(1,4) $ of the group $SO(1,14)$ determine spins of
fields, while generators
of the subgroups $ SU(3), SU(2), U(1) $ determine their charges.

The Lagrange function describing a particle on a supergeodesics,
leads to the momentum of the particle in Grassmann space which is
proportional to the Grassmann coordinate. This brings  the Clifford
algebra and the spinorial degrees of freedom into the theory.
The supervielbeins, transforming the geodesics from the freely
falling to the external coordinate system, carry the vectorial as
well as the spinorial degrees of freedom. All fields, the
fermionic and the bosonic fields, depend on ordinary and
Grassmann coordinates, the later determine spins and charges
of fields.

The Yang-Mills fields appear as the contribution of gravity
through spin connections and not through vielbeins as in the
Kaluza-Klein theories. Because of that and because
the generators of the Lorentz transformations in Grassmann
space  rather then momentum in
ordinary space determine charges of fields, the Planck mass of
charged particles and the transformation inconsistencies of
gauge fields in Kaluza-Klein theories do not appear.

More about this approach can be found in  Refs.\cite{man1,man2}.

\vskip 1cm

\noindent
{\bf COORDINATE GRASSMANN SPACE AND LINEAR OPERATORS }

\vskip 0.5cm

In this section we briefly repeat a few definitions concerning
a d-dimensional Grassmann space,  linear Grassmann space
spanned over the coordinate space, linear operators defined in
this space and the Lie algebra of generators of the Lorentz
transformations \cite{man2,ber}.

\vskip 0.5cm

\noindent
{\bf Coordinate space with Grassmann character}

\vskip 0.3cm

We define a d-dimensional Grassmann space of real anticommuting
coordinates $ \{ \theta ^a \} $, $ a=0,1,2,3,5,6,...,d,$
satisfying the anticommutation relations

$$ \theta^a \theta^b + \theta^b \theta^a := \{ \theta^a, \theta^b
\} = 0,
\eqno (2.1) $$

called the Grassmann algebra \cite{man2,ber}. The metric tensor $ \eta
_{ab}$ $ = diag (1,$ $ -1, -1,$ $ -1,..., -1) $ lowers the
indices of a
vector $\{ \theta^a \} = \{ \theta^0, \theta^1,..., \theta^d \},
\theta_a = \eta_{ab} \theta^b$. Linear transformation actions on
vectors $ (\alpha \theta^a + \beta  x^a) $

$$ (\alpha \acute{\theta}^a + \beta \acute{x}^a ) = L^a{ }_b
(\alpha \theta^b + \beta x^b ) ,\eqno (2.2) $$

which leave forms

$$ ( \alpha \theta^a + \beta x^a ) ( \alpha \theta ^b + \beta
x^b ) \eta_{ab} \eqno (2.3) $$

invariant, are called the Lorentz transformations. Here $
(\alpha \theta^a + \beta x^a ) $ is a  vector of d
anticommuting components (Eq.(2.1)) and d commuting $ (x^ax^b -
x^bx^a = 0) $ components, and $ \alpha $ and $ \beta$ are two
complex numbers. The requirement that forms (2.3) are scalars with
respect to the linear transformations (2.2), leads to the
equations

$$ L^a{ }_c L^b{ }_d \eta_{ab} = \eta_{cd}.  \eqno(2.4)$$

\vskip 0.5cm

\noindent
{\bf Linear vector space}

\vskip 0.3cm

A linear space spanned over a Grassmann coordinate space of d
coordinates has the dimension $ 2^d$. If monomials $
\theta^{\alpha_1} \theta^{\alpha_2}....\theta^{\alpha_n}, $
are taken as a set of basic vectors with $\alpha_i \neq
\alpha_j,$ half of the vectors have an even (those with an even n)
and half of the vectors have an odd (those with an odd n)
Grassmann character. Any vector in this space may be represented
as a linear superposition of monomials

$$ f(\theta) = \alpha_0 + \sum_{i=1}^{d}  \alpha _{a_1a_2 ..a_i}
\theta^{a_1} \theta^{a_2}....\theta^{a_i},\;\; a_k< a_{k+1},
\eqno(2.5)$$

where constants $\alpha_0, \alpha_{a_1a_2..a_i}$ are complex
numbers.

\vskip 0.5cm

\noindent
{\bf Linear operators}

\vskip 0.3cm

In  Grassmann space the left derivatives have to be
distinguished  from the right derivatives, due to the
anticommuting nature of the coordinates \cite{man2,ber}. We
shall make use
of left derivatives $ {\overrightarrow {{\partial}^{\theta}}}{
}_a:= \frac{\overrightarrow {\partial}}{\partial
\theta^a},\;\;\; {\overrightarrow {{\partial}^{\theta}}}{ }^{a}:=
{\eta ^{ab}} \overrightarrow {{\partial}^{\theta}}{ }_b \; $, on
vectors of the
linear space of monomials $ f(\theta)$,  defined as follows:

$$ {\overrightarrow{{\partial}^{\theta}}}{ }_a\; \theta^b f(\theta)
= \delta^b{ }_a f(\theta) - \theta^b
{\overrightarrow{{\partial}^{\theta}}}{ }_a\; f(\theta) , \eqno(2.6) $$

$$ {\overrightarrow {{\partial}^{\theta}}}{ }_a \alpha f(\theta) =
(-1)^{n_{a \partial}} \alpha {\overrightarrow{{\partial}^{\theta}}}{
}_a \;f(\theta).$$

Here  $ \alpha $ is a constant of either commuting $( \alpha
\theta^a - \theta^a \alpha = 0 )$ or anticommuting $( \alpha
\theta^a + \theta^a \alpha = 0 )$ character, and $n_{a
\partial}$ is defined as follows

$$ n_{AB} = \left\{ \begin{array} {ll} +1, & if\; A\; and\; B \;
have\;Grassmann\; odd\; character\\
0, & otherwise \end{array} \right\} $$

We define the following linear operators \cite{man1,man2}

$$ p^{\theta} { }_a := -i {\overrightarrow{{\partial}^{\theta}}}{
}_a , \;\;
 \tilde{a} ^a := i(p^{\theta a} - i \theta^a) ,\;\;
\tilde{\tilde{a}}{}^a := -(p^{\theta a} + i \theta^a). \eqno
(2.7) $$

According to the inner product defined in the next subsection,
the operators $  \tilde{a} ^a $ and $ \tilde{\tilde{a}}{}^a $
are either hermitian or antihermitian operators

$$ \tilde a^{a}{ }^+ = -\eta^{aa} \tilde a^a,\;\;\; \tilde{\tilde
a}{ }^{a}{ }^+ =  -\eta^{aa} \tilde{\tilde a}{ }^{a}. \eqno(2.7a)
$$

We define the generalized commutation relations \cite{man1,man2}
( we shall show later that they follow from the
corresponding Poisson brackets):

$$ \{ A,B \} := AB - (-1) ^{n_{AB}} BA, \eqno(2.8) $$

fulfilling the equations

$$ \{A,B\}  = (-1)^{n_{AB}+1} \{B,A\},                  \eqno(2.9a)$$

$$ \{A,BC \} = \{A,B\} C + (-1)^{n_{AB}}B \{A,C\},      \eqno(2.9b)$$

$$ \{AB,C\} = A\{B,C\} + (-1)^{n_{BC}}\{A,C\}B,
                                     \eqno(2.9c)$$

$$ (-1)^{n_{AC}}\{A,\{B,C\}\} + (-1)^{n_{CB}}\{C,\{A,B\}\}
+ (-1)^{n_{BA}} \{B,\{C,A\}\} = 0.                       \eqno(2.9d)$$

We find

$$ \{p^{\theta a}, p^{\theta b} \} = 0 = \{ \theta^{a},
\theta^{b}\}, \eqno(2.10a)$$

$$  \{p^{\theta a}, \theta^{b}\} = -i \eta^{ab},         \eqno(2.10b)$$
$$ \{\tilde{a}^{a}, \tilde{a}^{b} \} = 2 \eta^{ab}
= \{\tilde{\tilde{a}}{ }^{a}, \tilde{\tilde{a}}{ }^{b} \},      \eqno(2.10c)$$

$$ \{ \tilde{a}^{a}, \tilde{\tilde{a}}{ }^{b} \} = 0.       \eqno(2.10d)$$

We see that $\theta ^a $ and $ p^{\theta a} $ form a Grassmann
odd Heisenberg algebra, while $ \tilde a^a $ and $
\tilde{\tilde{a}}{ }^a $ form the Clifford algebra.

We  define the projectors

$$ P_{\pm} = \frac{1}{2} ( 1 \pm  \sqrt{ (-)^{\tilde
\Upsilon \tilde{\tilde
\Upsilon}}} \tilde{ \Upsilon} \tilde{\tilde \Upsilon}),\;\;\;\;
(P_{\pm})^2 =
P_{\pm}, \eqno (2.11) $$

where $\tilde \Upsilon$ and $ \tilde{\tilde \Upsilon}$ are the two
operators  defined  for any dimension d as follows

$$ \tilde \Upsilon = i^{\alpha} \prod_{a=0,1,2,3,5,..,d} \tilde{
a}{ }^a \sqrt{\eta^{aa}}, \eqno(2.11a) $$

$$ \tilde{\tilde \Upsilon} = i^{\alpha} \prod_{a=0,1,2,3,5,..,d}
\tilde{ \tilde{a}}{ }^a \sqrt{\eta^{aa}},\eqno(2.11b) $$

with $\alpha $ equal either to $ d/2 $ or to $ (d-1)/2 $ for
even and odd dimension $d$ of the space, respectively.

It can be checked that $( \tilde \Upsilon )^2 = 1 = ( \tilde{
\tilde{\Upsilon}} )^2 $.

The projectors $P_{\pm}$ project out of any monomials of
Eq.(2.5) the Grassmann odd and the Grassmann even part of the monomial,
respecticely.

We find that for odd d the operators $\tilde \Upsilon $ and
$\tilde{ \tilde \Upsilon}$ coincide ( up to $\pm i$ or $ \pm1$ )
with $\tilde
\Gamma$ and $\tilde{\tilde \Gamma}$ of Eq.(2.17), respectively.

\vskip 0.5cm

\noindent
{\bf Lie algebra of generators of Lorentz transformations}

\vskip 0.3cm

We define two kinds of operators \cite{man2}. The first ones are
binomials of operators forming the Grassmann odd Heisenberg
algebra

$$ S^{ab} : = ( \theta^a p^{\theta
b} - \theta ^b p^{\theta a} ).                     \eqno  (2.12a)$$

The second kind are binomials of operators forming the Clifford
algebra

$$ \tilde S ^{ab}: = - \frac{i}{4} [\tilde a ^a , \tilde a ^ b
], \;\; \tilde {\tilde S} { }^{ab}: = - \frac{i}{4} [ \tilde
{\tilde a}{ }^a , \tilde {\tilde a}{ }^b ] , \eqno(2.12b)$$

with $ [A, B]:= AB - BA.$

Either $ S^{ab} $ or $ \tilde S ^{ab}$  or $ \tilde{\tilde S}{
}^{ab} $ fulfil the Lie algebra of the Lorentz group $ SO(1,d-1)
$ in the d-dimensional Grassmann space :

$$ \{ M^{ab}, M^{cd} \} = -i ( M^{ad} \eta^{bc} + M^{bc}
\eta^{ab} - M^{ac} \eta^{bd} - M^{bd} \eta^{ac} ) \eqno(2.13)$$

with $ M^{ab} $ equal either to $ S ^{ab} $ or to $\tilde S
^{ab} $ or to $ \tilde {\tilde S} { }^{ab} $ and $ M^{ab} = -
M^{ba} $.

We see that

$$ S^{ab} = \tilde S ^{ab} + \tilde {\tilde S}{ }^{ab},\;\;
 \{ \tilde S ^{ab} , \tilde {
\tilde S }{ }^{cd} \} = 0 = \{ \tilde S ^{ab} , \tilde {\tilde
a}{ }^c \} = \{ \tilde a ^a , \tilde { \tilde S }{ }^{bc} \}.
\eqno (2.14)$$

By solving the eigenvalue problem (see Sect. 2.6) we find that
operators $ \tilde S ^{ab} $, as well as
the  operators $ \tilde {\tilde S}{ }^{ab} $, define the
fundamental
or the spinorial representations of the Lorentz group, while $
S^{ab} = \tilde S ^{ab} + \tilde{\tilde S} { }^{ab} $
define the regular or the adjoint or the vectorial
representations of the Lorentz group $ SO(1,d-1) $.

Group elements are in any of the three cases defined by:

$$ {\cal U}(\omega) = e^{ \frac{i}{2} \omega_{ab} M^{ab}}
,\eqno(2.15) $$

where $ \omega_{ab} $ are the parameters of the group.

Linear transformations, defined in Eq.(2.2), can then be written
in terms of group elements as follows

$ \acute{\theta}^a = L^a{ }_b \theta ^b = e^{- \frac{i}{2}
\omega_{cd} S^{cd}} \theta ^a e^{\frac{i}{2} \omega _{cd}
S^{cd}}. $

 By using Eqs.(2.9) and (2.13) it can be proved for any d , that
$ M^2 $ is the invariant of the Lorentz group

$$ \{ M^2, M^{cd} \} = 0,\;\; M^2 =  \frac{1}{2} M^{ab} M_{ab} ,
 \eqno(2.16) $$

and that for d=2n we can find the additional invariant $ \Gamma $

$$ \{ \Gamma, M^{cd} \} = 0,\;\; \Gamma = \frac{i(-2i)^{n}
}{(2n)!}
\epsilon_{a_1a_2...a_{2n}} M^{a_1a_2} ....M^{a_{2n-1}a_{2n}} ,
 \eqno(2.17) $$

where $\epsilon _{a_1a_2...a_{2n}} $ is the totally antisymmetric
tensor with $ 2n $ indices and with $ \epsilon _{ 1 2 3 ...2n }
= 1 $. This means that $ M^2  $
and $ \Gamma $ are for $ d = 2n$ the two invariants or Casimir
operators of the group $ SO(d) $ (or $ SO(1,d-1) $, the
two algebras differ
only in the definition of the metric $ \eta^{ab} $).
For  $ d = 2n + 1 $ the
second invariant cannot be defined.( It can be checked that
$\tilde \Upsilon$ and
$\tilde{\tilde \Upsilon} $ of eqs.(2.11) are the two invariants
for the spinorial case for any d. For  even d they coincide with
$\tilde \Gamma $ and $\tilde{\tilde \Gamma}$, respectively,
while for odd d
the eigenvectors of these two operators are superpositions of
Grassmann odd and Grassmann even monomials.)

While the invariant $ M^2 $ is trivial in the
case when $ M^{ab} $ has spinorial character, since
 $ (\tilde S^{ab})^2 = \frac{1}{4} \eta^{aa} \eta^{bb} =
(\tilde {\tilde S}{ }^{ab})^2 $ and therefore $ M^2 $ is equal
in both cases  to the number $ \frac{1}{2} \tilde
S^{ab} \tilde S _{ab} = \frac{1}{2} \tilde {\tilde S}{ }^{ab}
\tilde {\tilde
S}{ }_{ab} = d ( d-1 ) \frac{1}{8}$ , it is a nontrivial
differential operator in  Grassmann space if $ M^{ab}$
have  vectorial character $(M^{ab} = S^{ab})$. The invariant of
Eq.(2.17) is always a nontrivial operator.

\vskip 0.5cm

\noindent
{\bf Integrals on Grassmann space. Inner products}

\vskip 0.3cm

We assume that differentials of Grassmann coordinates $ d\theta^a
$ fulfill the Grassmann anticommuting relations \cite{man2,ber}

$$ \{ d\theta^a, d\theta^b \} = 0  \eqno(2.18)$$

and we introduce a single integral over the whole interval of
$d\theta^a $

$$ \int d\theta^a = 0, \;\; \int d\theta^a \theta^a = 1, a =
0,1,2,3,5,..,d, \eqno(2.19)$$

and the multiple integral over d coordinates

$$ \int d^d \theta^0 \theta^1 \theta^2 \theta^3
\theta^4...\theta^d = 1, \eqno(2.20) $$

with

$$ d^d \theta: = d\theta^d...d\theta^3 d\theta^2 d\theta^1
d\theta^0. $$

We define \cite{man2,ber} the inner product of two vectors $
<\varphi|\theta> $ and $ <\theta|\chi> $, with $
<\varphi|\theta> = <\theta|\varphi>^* $ as follows:

$$ <\varphi|\chi> = \int d^d\theta ( \omega<\varphi|\theta>)
<\theta| \chi>, \eqno(2.21) $$

with the weight function $\omega$

$$\omega = \prod_{k=0,1,2,3,..,d}
(\frac{\partial}{\partial \theta^k}  + \theta^k ),
\eqno(2.21a)$$

which operates on the first function $ <\varphi|\theta> $ only,
and we define

$$ (\alpha^{a_1 a_2...a_k} \theta^{a_1}
\theta^{a_2}...\theta^{a_k})^{+} =
(\theta^{a_k}).....(\theta^{a_2})
(\theta^{a_1}) (\alpha^{a_1 a_2...a_k})^{*}.\eqno(2.21b)$$

According to the above definition of the inner product it
follows that  $\tilde a^a{ }^+ = - \eta^{aa} \tilde a^a $ and
$\tilde{\tilde a}{ }^a{ }^+ = - \eta^{aa} \tilde{\tilde a}{ }^a
$, $(\tilde a^a \tilde a^b){ }^+ = - \eta^{aa} \eta^{bb}
\tilde a^a \tilde a^b $, and $(\tilde{\tilde a}{ }^a
\tilde {\tilde a}{ }^b){ }^+ = - \eta^{aa} \eta^{bb}
\tilde {\tilde a}{ }^a \tilde {\tilde a}{ }^b. $  The
generators of the Lorentz
transformations (eqs.(2.12a) and (2.12b)) are self adjoint ( if $
a \neq 0 $ and  $ b \neq 0 $ ) or anti self adjoint ( if $ a =
0 $ or $ b = 0 $ )  operators.


According to eqs.(2.7) and (2.12) we find

$$ S^{ab} = -i ( \theta^a \frac{\partial}{\partial \theta_b}
- \theta^b \frac{\partial}{\partial \theta_a} ), \eqno(2.22a) $$

$$ \tilde a^a = (\frac{\partial}{\partial \theta_a} +
\theta^a ),\;\; \tilde{\tilde a}{ }^a = i
(\frac{\partial}{\partial \theta_a} -
\theta^a ). \eqno(2.22b) $$

$$ \tilde S ^{ab} = \frac{-i}{2}( \frac{\partial}{\partial
\theta_a} +
\theta^a ) ( \frac{\partial}{\partial \theta_b} +
\theta^b ) ,\;\;  \tilde {\tilde S}{ } ^{ab} = \frac{i}{2}(
\frac{\partial}{\partial \theta_a} -
\theta^a ) ( \frac{\partial}{\partial \theta_b} -
\theta^b ) ,\; if a \neq b, \; \eqno(2.22c) $$

 \vskip 0.5cm

\noindent
{\bf The eigenvalue problem}

\vskip 0.3cm

To find  eigenvectors of any operator $A$, we
solve the eigenvalue problem

$$ <\theta|\tilde A_i|\tilde{\varphi}> = \tilde{\alpha}_i
<\theta|\tilde{\varphi}> ,\;\;
<\theta|A_i|\varphi> = \alpha_i <\theta|\varphi>,\;\;i = \{1,r\}
,\eqno(2.23)$$

where $ \tilde{A}_i $ and $ A_i $ stand for $r$ commuting
operators
of spinorial and vectorial character, respectively.

To solve  equations (2.23) we express the operators in the
coordinate representation (Eqs.(2.22)) and write the eigenvectors
as polynomials of $\theta^a$. We  orthonormalize the vectors
according to the inner product, defined in Eq.(2.21)

$$ <{ }^a \tilde{\varphi}_i |{ }^b \tilde{\varphi}_j> =
\delta^{ab} \delta_{ij},\;\;\; <{ }^a {\varphi}_i |{ }^b
{\varphi}_j> = \delta^{ab} \delta_{ij}, \eqno(2.23a)$$

where index $a$ distinguishes between vectors of different
irreducible representations and index j between vectors of the
same irreducible representation. Eq.(2.23a) determines the
orthonormalization condition for spinorial and vectorial
representations, respectively.

\vskip 1cm

\noindent
{\bf LORENTZ GROUPS AND SUBGROUPS }

\vskip 0.5cm

 The algebra of the group $ SO(1,d-1) $ \hbox{\it or} $ SO(d) $
contains \cite{man1}  $ n $ subalgebras defined by
operators $ \tau ^{A i}, A = 1,n ; i = 1,n_A $,
where $ n_A $ is the number of elements of each
subalgebra, with the properties

$$ [ \tau ^{Ai} , \tau ^{B j} ] = i \delta ^{AB
} f^{A ijk } \tau ^{A k}, \eqno (3.1) $$

if operators $ \tau ^{A i} $  can be expressed as
linear superpositions  of operators $ M^{ab} $

$$ \tau^{A i} = \hbox{\it c} ^{A i} { }_{ab} M^{ab}, \;\;
\hbox{\it c} ^{A i}{ }_{ab} = - \hbox{\it c} ^{A i}{ }_{ba}, \;\;
A=1,n, \;\;
i=1,n_{A}, \;\;a,b=1,d. \eqno(3.1a) $$

Here $ f^{A ijk} $ are structure constants
of the ($ A $) subgroup with $ n_{A} $ operators.
According to the three kinds of operators $ M^{ab} $, two of
spinorial and one of vectorial character, there are  three kinds
of operators $ \tau^{A i} $ defining subalgebras of
spinorial and vectorial character, respectively, those of
spinorial types being expressed with either $ \tilde S^{ab} $ or
$ \tilde{ \tilde S}{ }^{ab} $ and those of vectorial type being
expressed by $ S^{ab} $. All three kinds
of operators are, according to Eq.(3.1),
defined by the same coefficients $ \hbox{\it c}^{A i} { }_{ab} $
and the same structure constants $ f^{A i j k } $.
{}From Eq.(3.1) the following relations among constants $\hbox{\it
c}^{A i}{ }_{ab} $ follow:

$$ -4 \hbox{\it c}^{A i}{ }_{ab} \hbox{\it c}^{B j b}{ }_c -
\delta^{A B} f^{A ijk} \hbox{\it c}^{A k}{ }_{ac}
= 0. \eqno (3.1b)$$

In the case when the algebra and the chosen subalgebras are
isomorphic, that is if the number of  generators of
subalgebras is equal to $ \frac{d(d-1)}{2} $ , the inverse
matrix $ e^{A iab} $ to the matrix of coefficients $
c^{A i}{ }_{ab} $ exists \cite{man1}
$M^{ab} = \sum_{A i} e^{A iab} \tau^{A i},$
with the properties $ c^{A i}{
}_{ab} e^{B jab} = \delta^{A B} \delta^{ij}, \;\;
c^{A i}{ }_{cd} e^{A iab} = \delta^a{ }_c \delta^b{
}_d - \delta^b{ }_c \delta^a{ }_d $.

When we look for coefficients $ c^{A i}{ }_{ab} $ which
express operators $ \tau ^{A i} $, forming a subalgebra
$ SU(n) $ of an algebra $ SO(2n) $ in terms of $ M^{ab} $, the
procedure is rather simple \cite{geor,man2}. We
find:

$$ \tau^{A m} = -\frac{i}{2} (\tilde \sigma^{A m})_{jk}
 \{ M^{(2j-1) (2k-1)} +
M^{(2j) (2k)} + i M^{(2j) (2k-1)} - i M^{(2j-1) (2k)}
\}.\eqno(3.2)$$

Here $(\tilde \sigma^{A m})_{jk}$ are the traceless matrices
which form the algebra of $ SU(n) $.
One can easily prove that operators $ \tau^{A m} $ fulfil the
algebra of the group $ SU(n) $  for any of three
choices for operators $ M^{ab} : S^{ab}, \tilde S^{ab},
\tilde{\tilde S}{ }^{ab}$.

In reference \cite{man2}  coefficients $\hbox{\it
c}^{A i}{ }_{ab} $ for
a few cases  interesting for  particle
physics can be found. Of special interest is the group $SO(1,14)$ with
the subgroups  $ SO(1,4)$ and $SO(10) \supset SU(3)\times SU(2)
\times U(1)$ which enables the unification of spins and charges.
As have we already said, the coefficients are the
same for all three kinds of operators, two of spinorial and one of
vectorial character.  The representations, of course, depend on the
operators $M^{ab}$ (See Eqs.(2.22)). After solving the
eigenvalue problem (Eqs.(2.23)) for the invariants of the
subgroups, the representations can be presented as polynomials
of coordinates $\theta^a, a = 0,1,2,3,5,..,15 $. The operators
of spinorial character define the fundamental representations of
the group and the subgroups, while the operators of vectorial
character define the adjoint representations of the groups.

We shall comment on the representations in the last section.

\vskip 1cm

\noindent
{\bf LAGRANGE FUNCTION FOR  FREE PARTICLES IN ORDINARY
AND GRASSMANN SPACE AND CANONICAL QUANTIZATION}

\vskip 0.5cm

We  present in this section the Lagrange function for a
particle which lives in a d-dimensional ordinary space
of commuting coordinates and in a d-dimansional Grassmann space
of anticommuting coordinates $ X^a \equiv \{ x^a, \theta^a \} $
and has its geodesics parametrized by an ordinary Grassmann even
parameter ($\tau$) and a Grassmann odd parameter($\xi$).  We
derive the Hamilton function and the corresponding Poisson
brackets and perform
the canonical quantization, which leads to the Dirac equation with
operators, which are differential operators in
ordinary and in  Grassmann space.

$ X^a = X^a(x^a,\theta^a,\tau,\xi)$ are called supercoordinates.
We define the dynamics of a particle by choosing the
action \cite{man1,ikem}

$$ I= \frac{1}{2} \int d \tau d \xi E E^i_A \partial_i X^a E^j_B
\partial_j X^b  \eta_{a b} \eta ^{A B} , \eqno (4.1) $$

where $ \partial _i : = ( \partial _ \tau , {\overrightarrow
{\partial}} _\xi ), \tau^i = (\tau, \xi) $, while $ E^i _A
$ determines a metric on a two dimensional superspace $ \tau ^i
$ , $ E = det( E^i _A )$ . We choose $ \eta _{A A} = 0,
\eta_{1 2} = 1 = \eta_{2 1} $, while $ \eta_{a b} $ is the
Minkowski metric with the diagonal elements $
(1,-1,-1,-1,$ $...,-1) $. The action is invariant under the Lorentz
transformations of supercoordinates: $X'{ }^a = L^A{ }_b X^b $.
(See Eq.(2.3)).
Since a supermatrix $ E^i{ }_A $ transforms as a vector in a
two-dimensional superspace $\tau^i$ under general coordinate
transformations of $\tau^i$, $ E^i{ }_A \tau_i $ is invariant
under such transformations and so is $d^2 \tau E$. The action
(4.1) is therefore locally supersymmetric. The inverse matrix $
E^A{ }_i$
is defined as follows: $E^i{ }_A E^B{ }_i = \delta^B{ }_A$.

Taking into account that either $ x^a $ or $ \theta^a $ depend
on an ordinary time parameter $ \tau $ and that $ \xi^2 = 0 $ ,
the geodesics can be described  as a
polynomial of  $ \xi $ as follows:
$ X^a = x^a + \varepsilon \xi \theta^a $. We choose $
\varepsilon^2 $
to be equal either to $ +i $ or to $ -i $ so that it defines two
possible combinations of supercoordinates. Accordingly we
also choose  the metric  $ E^i { }_A $ : $ E^1{ }_1 = 1, E^1{
}_2 = - \varepsilon M, E^2{ }_1 = \xi, E^2{ }_2 = N -
\varepsilon \xi M $, with $ N $ and $ M $  Grassmann even and
odd parameters, respectively. We write $ \dot{A} =
\frac{d}{d\tau}A $, for any $ A $.

If we integrate the action(4.1) over the Grassmann odd
coordinate $d\xi$, the action for a superparticle follows:

$$ \int d\tau ( \frac{1}{N} \dot{x}^a \dot{x}_a + \varepsilon^2
\dot{\theta}^a \theta_a - \frac{2\varepsilon^2 M}{N} \dot{x}^a
\theta_a ). \eqno(4.1a)$$

Defining the two momenta

$$ p^{\theta }_a : = \frac{ \overrightarrow{\partial} L}
{ \partial {\dot{\theta}^a}} = \epsilon^2 \theta^a , \eqno(4.2a)$$.

$$ p_a : = \frac{\partial L}{ \partial \dot{x}^a} = \frac{2}{N}(
\dot{x}_a - M p^{\theta a}), \eqno(4.2b)$$

the two Euler-Lagrange equations follow:

$$ \frac{dp^a}{d \tau} = 0,\;\;\; \frac{dp^{\theta a}}{d \tau} =
\varepsilon ^2 \frac{M}{2} p^a. \eqno(4.3)$$

Variation of the action(4.1a) with respect to $M $ and $N$ gives
the two constraints
$$ \chi^1: = p^a a^{\theta}_a = 0, \chi^2 = p^a p_a = 0, \;\;
a^{\theta}_a:= i p^{\theta}_a + \varepsilon^2 \theta_a,
\eqno(4.4)$$

while
$ \chi^3{ }_a: = - p^{\theta }_a + \epsilon^2 \theta_a = 0 $
(eq.(4.2a) is the third type of constraints of the action(4.1).
For $\varepsilon^2 = -i$ we find, if using Eq.(2.7), that $
a^{\theta}{ }_a = \tilde{a}
^a,\;\; \chi^3{ }_a = \tilde{\tilde a}_a = 0. $

We find the generators  of the Lorentz transformations for the
action(4.1) to be (See Eq.(2.12))

$$ M^{ a b} = L^{a b} + S^{a b} \;,\; L^{a b} = x^a p^b - x^b p^a
\;,\; S^{a b} = \theta^a p^{ \theta b} - \theta^b p^{ \theta a}
=  \tilde{S} ^{a b} + \tilde{\tilde{S}}{}^{a b},
\eqno (4.5) $$

which show that parameters of the Lorentz transformations are the
same in both spaces.

We define the Hamilton function:

$$ H:= \dot{x}^a p_a + \dot{\theta}{ }^a p^{\theta}{ }_a - L =
\frac{1}{4} N p^a p_a + \frac{1}{2} M p^a (\tilde a_a +
i \tilde{\tilde a }{ }_a) \eqno(4.6)$$

and the corresponding Poisson brackets

$$\{A,B\}_p=
\frac{ \partial A}{ \partial x^a} \frac{ \partial B}{ \partial
p_a}  - \frac{ \partial A}{ \partial p_a} \frac{ \partial B}{ \partial
x^a} +  \frac{ \overrightarrow{ \partial A}}{\partial \theta
^a} \frac{ \overrightarrow{ \partial B}}{\partial p^\theta_a} +
  \frac{ \overrightarrow{ \partial A}}{\partial p^\theta_a}
\frac{ \overrightarrow{ \partial B}}{\partial \theta^a}, \eqno
(4.7) $$

which have the properties of the generalized commutators
presented in Eqs.(2.8-2.9).

If we take into account the constraint $\chi^3{ }_a = \tilde{\tilde
a}{ }_a = 0\;$ in the Hamilton function (which just means that
instead of H the Hamilton function $ H + \sum_i \alpha^i
\chi^i + \sum_a \alpha^3{ }_a \chi^3{ }^a $ is taken, with parameters $
\alpha^i, i=1,2 $ and $ \alpha^3{ }_a = -\frac{M}{2} p_a,
a=0,1,2,3,5,..,d $
chosen on ssuch a way that the Poisson brackets
of the three types of constraints with the new Hamilton function
are equal to zero) and in all dynamical quantities, we find:

$$ H = \frac{1}{4} N p^a p_a + \frac{1}{2} M p^a \tilde a_a,\;\;
\chi^1 = p^a p_a = 0,\;\; \chi^2 = p^a \tilde a_a =
0,\eqno(4.4a) $$

$$ \dot{p}_a = \{ p_a, H \}_P = 0, \dot{\tilde a}{ }_a = \{
\tilde{a}_a, H \}_P = iM p_a,\eqno(4.3a) $$

which agrees with the Euler Lagrange equations (4.3).

We further find
$$ \dot {\chi}^i = \{ H, \chi^i \}_P = 0,\;\;i =1,2,\;\;\; \dot
{\chi}^3{ }_a = \{ H, \chi^3{ }_a \}_P = 0,\;\;a =
0,1,2,3,5,..,d,\eqno(4.3b) $$

which guarantees that the three
constraints will not change with the time parameter $\tau$ and
that $\dot{\tilde M}{ }^{ab} = 0 $, with $ \tilde M { }^{ab} =
L^{ab} + \tilde{S}^{ab}$,  saying that $ \tilde M{ }^{ab} $
is the constant of motion.

The Dirac brackets, which can be obtained from the Poisson
brackets of Eq.(4.7) by adding to these brackets on the right
hand side a term $ - \{A, \tilde{\tilde a}^c \}_P \cdot$ $ ( -
\frac{1}{2i} \eta_{ce} ) \cdot $ $ \{ \tilde{\tilde a}{ }^e, B
\}_P $, give  for the dynamical quantities,
which are observables, the same results as the Poisson bracket.
This is true also for $ \tilde a^a,$ ( $\{ \tilde
a^a, \tilde a^b \}_D = i\eta^{ab} = \{
\tilde a^a, \tilde a^b \}_P $),  which is the
dynamical quantity but not  an observable since its odd
Grassmann character causes  supersymmetric
transformations. We also find that $\{ \tilde a^a, \tilde{\tilde
a}{ }^b \}_D
= 0 = \{ \tilde a^a, \tilde{\tilde a}{ }^b \}_P $ .
The Dirac brackets give  different results only for the quantities
$ \theta^a $ and $ p^{\theta a} $ and  for $\tilde {\tilde
a}{ }^a $ among themselves: $ \{ \theta^a, p^{\theta b}
\}_P = \eta^{ab}, \{ \theta^a, p^{\theta b}
\}_D = \frac{1}{2} \eta^{ab} $, $ \{ \tilde {\tilde a}{ }^a, \tilde
{\tilde a}{ }^b \}_P = 2i \eta^{ab}, \{ \tilde {\tilde a}{ }^a,
\tilde {\tilde a}{ }^b \}_D = 0 $. According to the above  properties
of the Poisson brackets, I suggest that in the quantization
procedure the Poisson brackets (4.7) rather then the Dirac
brackets are used, so that variables $\tilde{\tilde a}^a $,
which are remooved from all dynamical quantities, stay as
operators. Then $\tilde a^a $ and
$\tilde{\tilde a}{ }^a $ are expressible with $\theta^a $ and
$p^{\theta a} $ (Eq.(2.7) and the algebra of linear operators
introduced in Sect.2 (Eqs.(2.7) - (2.14)) can be used. We shall
show, that  suggested quantization procedure leads to the
Dirac equation, which is the differential equation in ordinary
and Grassmann space and has all desired properties.

In the proposed quantization procedure$ -i \{ A,B \}_p $ goes to
either a commutator or to an anticommutator, according to the
Poisson brackets (4.7). The operators $\theta ^a , p^{\theta a}
$ ( in the coordinate representation they become $ \theta^a
\longrightarrow \theta^a , \; p^{\theta}_a \longrightarrow i
\frac{\overrightarrow{\partial }}{\partial \theta^a} $) fulfill
the Grassmann odd Heisenberg algebra, while the operators
$ \; \tilde{a}^a \; $ and $\; \tilde{\tilde{a}}{}^{a}\; $ fulfill
the Clifford algebra(Eqs.(2.10))

The constraints (Eqs.(4.3)) lead to
the Dirac like and  the Klein-Gordon equations

$$ p^a \tilde{a} _a | \tilde{\Psi} > = 0 \;,\; p^a p_a |
\tilde{\Psi}> = 0 , \; with \;  p^a \tilde{a}_a p^b \tilde{a}_b =
p^a p_a . \eqno  (4.8) $$

Trying to solve the eigenvalue problem $ \tilde{\tilde a}{ }^a
| \tilde {\Psi} > = 0,\;\; a=(0,1,2,3,5,...,d), $ we find that no
solution of this eigenvalue problem exists, which means that
the third constraint $ \tilde{\tilde a}{ }^a = 0 $ can't be
fulfilled in the operator form (although we take it into account
in the operators for all dynamical variables in order that
operator equations would agree with classical equations). We can
only take it into account
in the  expectation value form

$$ < \tilde{\Psi} | \tilde{\tilde a}{ }^a | \tilde{\Psi} > = 0.
\eqno(4.9) $$

Since $ \tilde{\tilde a}{ }^a $ are Grassmann odd operators,
they change monomials (Eq.(2.5)) of an Grassmann odd character
into monomials of an Grassmann even character and opposite,
which is the supersymmetry transformation.
It means that Eq.(4.9) is fulfilled for momomials of either odd
or even Grassmann character and that superpositions of the
Grassmann odd and the Grassmann even monomials are not solutions
for this system.

We can use the projector $P_{\pm}$ of Eq.(2.11) to project
out of monomials  either the Grassmann odd or the Grassmann even
part. Since this projector commutes with the Hamilton function $
( H =
\frac{N}{4} p^a p_a + \frac{1}{2} \mu\; p^a \tilde a_b \tilde
a_a,\;\; \{ P_{\pm}, H \} = 0 ) $, where we choose $M$ to be
proportional to $\tilde a_b $, for any $ b\subset \{0,1,..,d\}$
and $ \mu $ is a real parameter, it means that eigenfunctions of $
H $ have either an odd or an even Grassmann character.
In order that in the second quantization procedure  fields
$ | \tilde{\Psi} > $ would describe fermions, it is meaningful
to accept  in the fermion case Grassmann  odd monomials only.

We further see that although the operators $ \tilde{a}^a $ fulfil
Clifford algebra, they cannot be recognized
as the Dirac $ \tilde{\gamma}^a $ operator, since they have an
odd Grassmann
character and therefore transform fermions into bosons, which is
not the case with the Dirac $ \gamma^a $ matrices. We
therefore recognize the generators of the Lorentz
transformations $ -2i \tilde{S}^{b m},\; m = 0,1,2,3 $, with $ b = 5
$ as the Dirac $ \gamma ^m $ operators.

$$\tilde{\gamma} ^m = - \tilde{a} ^5 \tilde{a} ^m = -2i\tilde{S}
^{5m} \;,\; m=0,1,2,3.  \eqno (4.10)$$

We choose  the Dirac operators $\tilde{\gamma}^a$ in
the way which in the case that  $ <\tilde{\psi}|p^{5}|
\tilde{\psi} >= m $ and $ < \tilde{\psi}|p^{h}|\tilde{\psi} > = 0
$,  for $ h \in \{ 6,d \} $, enables to recognize the equation

$$ (\tilde{\gamma}{ }^m p_{m} - m )|\tilde{\psi}> = 0 \;,\;
m=0,1,2,3. \eqno  (4.11) $$

as the Dirac equation. The $m = 5$ is just the number of the
previledged coordinate which determines the operators $ \tilde
{\gamma} ^a $.
Since $-2i\tilde{S}^{5m}$ appear as $ \tilde{\gamma}{ }^m, $
$ SO(1,4) $ rather than $ SO(1,3)$ is needed
to describe the spin degrees of freedom of fermionic fields.

It can be checked that in the
four-dimensional subspace $\tilde{\gamma}{ }^{m} $ fulfill
the Clifford algebra $\{\tilde{\gamma}{ }^{m} , \tilde{\gamma}{
}^{n}\}  = \eta{^{mn}} $ , while $ \tilde{S}{ }^{mn} = -\frac{i}{4}
\lbrack \tilde{\gamma}{ }^{m},\tilde{\gamma}{ }^{n}\rbrack_{-}
$.

We presented in  Ref. \cite{man2} four Dirac four spinors ( the
polynomials of $ \theta^a $) which fulfil  Eq.(4.11) if $ m \ne
0 $ and four Weyl four spinors which fulfill  Eq.(4.11) if $
m= 0. $

For large enough d not only do  generators of  Lorentz
transformations (of a spinorial character ) in  Grassmann
space define
the spinorial degrees of freedom of a particle field in the four
dimensional subspace, but they also define the quantum numbers of
these fields, which may be recognized
as electromagnetic, weak and colour charges.

This certainly can be done for $d = 15 $, since $ SO(1,14) $ has
the subalgebra $ SO(1,4)\times SO(10) $, while $ SO(10) $ has
the subalgebra $ SU(3)\times SU(2)\times U(1) $.
In this case  $\tilde{ \tau}^{A i}$ are  linear
superpositions of operators $ \tilde S ^{ab},\; a,b \in \{6,d\}$
fulfilling the algebras as presented in Eqs.(3.1-3.3)

$$ [\tilde{ \tau}^{A i} , \tilde{ \tau}^{B j}] = i
\delta^{AB} f^{A ijk} \tilde{ \tau}^{A k}, \hbox{\rm for}\;\;
A, B,  \eqno(3.1a) $$

and defining the algebras of $ SU(3), SU(2), U(1),$
while $ SO(1,4) $ remains to define the spinorial degrees of
freedom in the four dimensional subspace. We find the fundamental
representations of the corresponding Casimir operators as
functions of $ \theta^a $ determining isospin doublets, colour
triplets and electromagnetic singlets \cite{man2}.

\vskip 1cm

\noindent
{\bf PARTICLES IN GAUGE FIELDS}

\vskip 0.5cm

The dynamics of a point particle in gauge fields, the
gravitational and the Yang-Mills fields, can be obtained by
transforming  vectors from a freely falling to an external
coordinate system \cite{wess}.
To do this, super vielbeins $\hbox{\bf
e}^{ia}{ } _{\mu} $ have to be
introduced, which in our case depend on ordinary and on
Grassmann coordinates, as well as on
two types of parameters $ \tau^i = ( \tau, \xi ) $. Since there
are two kinds of derivatives $ \partial_i $, there are two
kinds of vielbeins \cite{man1,man2}. The index a
refers to a freely falling coordinate system ( a Lorentz index),
the index $\mu$ refers to an external coordinate system ( an
Einstein index). Vielbeins with a Lorentz index smaller than
five will determine ordinary gravitational fields, and those with
a Lorentz index higher than four will define Yang-Mills
fields. Spin connections appear in the theory as ( a part of)
Grassmann odd fields.

We write the transformation of vectors as follows

$$ \partial_i X^a= \hbox{\bf e}^{i a} { }_{\mu} \partial_i X^{\mu} \;,\;
\partial_i X^{\mu} = \hbox{\bf f}^{i \mu} { }_a \partial_i X^a \;,\;
\partial_i = ( \partial_{\tau} , \partial_{\xi} ) . \eqno   (5.1)$$

 From eq.(5.1) it follows that

$$ \hbox{\bf e}^{i a} { }_{\mu} \hbox{\bf f}^{i \mu} { }_b =
\delta^a { }_b \;,\;
\hbox{\bf f}^{i \mu} { }_{a} \hbox{\bf e}^{i a} { }_{\nu} = \delta^{\mu} {
}_{\nu} .\eqno (5.2) $$

Again we make a Taylor expansion of vielbeins with respect to
$ \xi $

$$ \hbox{\bf e}^{i a} { }_{\mu} = e^{i a} { }_{\mu} + \varepsilon \xi
\theta^b e^{i a} { }_{ \mu b} \;,\; \hbox{\bf f}^{i \mu} { }_a = f^{i
\mu} { }_a - \varepsilon \xi \theta^b
f^{i \mu} { }_{a b} \;,\;i=1,2.  \eqno (5.3) $$

Both expansion coefficients  again depend  on ordinary
and on Grassmann coordinates. Since $ e^{ia} { }_{\mu}$ have an
even Grassmann character they will describe the spin 2 part of a
gravitational field. The coefficients $ \varepsilon \theta^{b}
e^{ia} { }_{\mu b}$ have an odd Grassmann
character ($\varepsilon$ is again the complex constant,
we choose $ \varepsilon^2$ equal to $-i $, so that
$\tilde{\tilde a} ^a = 0 $). We shall see that they define the
spin connections \cite{man1,man2}.

{}From Eqs.(5.2) and (5.3) it follows that

$$   e^{i a} { }_{\mu} f^{i \mu} { }_b = \delta^a { }_b \;,\;
f^{i \mu} { }_{a} e^{i a} { }_{\nu} = \delta^{\mu} { }_{\nu}
\;,\; e^{i a} { }_{\mu b} f^{i \mu} { }_c = e^{i a} { }_{\mu}
f^{i \mu} { }_{c b} \;,\; i=1,2. \eqno (5.2a) $$

We find the metric tensor $\hbox{\bf g}^{i}_{\mu \nu} =
\hbox{\bf e}^{ia}
{ }_{\mu} \hbox{\bf e}^{i}_{a \nu} ,\;
\hbox{\bf g}^{i \mu \nu} =\hbox{\bf f}^{i \mu} { }_{a}
\hbox{\bf f}^{i \nu a} ,
i=1,2$,  with an even Grassmann character and the properties
$\hbox{\bf g}^{i \mu \sigma}
\hbox{\bf g}^{i}_{\sigma \nu} = \delta ^{\mu}
 { }_{\nu}= g^{i \mu \sigma} g ^{i}_{\sigma \nu}$, with
$g^{i}_{\mu \sigma} = e^{ia} { }_{\mu} e^{i} { }_{a \sigma} $.

It follows from Eq.(5.1) that vectors in a freely falling and in
an external coordinate system are connected as follows:
$ \dot{x}^a= e^{1 a} { }_{\mu} \dot{x}^{\mu} \;,\; \dot{x}^{\mu}
= f^{1 \mu} { }_a \dot{x}^a\;,\; \theta^a=e^{2 a} { }_{\mu}
\theta^{\mu} \;,\; \theta_{\mu} = f^{2 \mu} { }_a \theta^a, $ and
$ \dot{\theta}^a= e^{1 a} { }_{\mu} \dot{\theta}^{\mu} + \theta^b
e^{1 a} { }_{\mu b} \dot{x}^{\mu} = (e^{2 a} { }_{\mu}
\theta^{\mu})^. =  e^{2 a} { }_{\nu , \mu_x} \dot{x}^\mu
\theta^{\nu} + e^{2 a} { }_{\mu} \dot{\theta}^{\mu} +
\dot{\theta}^{\mu}
\overrightarrow{e^{2 a}} { }_{\nu ,\mu_{\theta} } \theta^{\nu}. $

We use the notation $e^{2a} { }_{\nu,\mu^{x}} = \frac{\partial}{
\partial x{^\mu}} e^{2a} { }_{\nu},
\overrightarrow{e^{2a}} { }_{\nu,\mu^{\theta}} =
\frac{\overrightarrow{\partial}}{\partial \theta^{\mu}}
e^{2a} { }_{\nu}$ .

The above equations define the following relations among the
fields:

$$ e^{ 2 a} { }_{\mu b}=0 \;,\; \overrightarrow{e^{2 a}} {
}_{\nu , \mu^{\theta}} \theta^{\nu} =
e^{1 a} { }_{\mu} - e^{2 a} { }_{\mu} \;,\;
 e^{1 a} { }_{\mu b} = e^{2 a} { }_{\nu , \mu^x} f^{2 \nu} {
}_b, \eqno (5.4)$$

which means that a point particle with a spin sees a spin
connection $ \theta^{b} e^{ia} { }_{\mu b} $ related to a
vielbein $ e^{2a} { }_{\nu}$.

Rewriting the action(4.1) in terms of an external coordinate
system according to Eqs.(5.1), using the Taylor expansion of
supercoordinates $ X^{\mu}$ and superfields $ \hbox{\bf{e}}^{ia} {
}_{\mu}$ and
integrating the action over the Grassmann odd parameter $\xi$,
the action

$$ I=\int d\tau \{ \frac{1}{N} g^1_{\mu \nu} \dot{x}^\mu
\dot{x}^\nu \; - \; \epsilon^2 \frac{ 2 M}{N} \theta_a e^{1 a} {
}_{\mu} \dot{x}^\mu \; + \; \varepsilon^2 \frac{1}{2}(
\dot{\theta}^\mu \theta_a -\theta_a \dot{\theta}^\mu) e^{1 a} {
}_{\mu} \; + $$
$$ + \;  \varepsilon^2 \frac{1}{2} (\theta^b \theta_a
-\theta_{a} \theta^b ) e^{1 a} { }_{  \mu b} \dot{x}^\mu \} ,$$
$$\eqno  (5.5) $$

defines the two momenta of the system

$$ p_{\mu} = \frac{\partial L}{\partial \dot{x}^\mu} = p_{0 \mu} +
 \frac{1}{2} \tilde{S}^{ab} e^1_{a \mu b} , \;\;
 p^\theta_\mu = -i \theta_a e^{1 a} { }_{\mu} = -i
(\theta_\mu + \overrightarrow{e}^{2 a} { }_{\nu , \mu_{\theta}}
e^2 { }_{a \alpha} \theta^{\nu} \theta^{\alpha}) , \eqno (5.6)$$

( where we made a choice of $\varepsilon^2 = -i $ ).
Here $ p_{0 \mu} $ are the covariant ( canonical) momenta of a
particle.
For $ p^{\theta}_{a} = p^{\theta}_{\mu} f^{1 \mu} { }_{a}$ it follows
that $ p^{\theta}_{a}$ is proportional to $\theta_{a}$. Then $
\tilde{a}_{a} = i
(p^{\theta}_{a} - i \theta_{a}),
$ while $ \tilde{\tilde{a}}_{a}= 0 $. We took this into account in
the left hand expression of Eq.(5.6). We may further write

$$ p_{ 0 \mu} = p_{ \mu} - \frac{1}{2} \tilde{S}^{a b} e^1_{a \mu b}
= p_{ \mu} - \frac{1}{2} \tilde{S}^{a b} \omega_{a b \mu} \;,\;
\omega_{a b \mu}=\frac{1}{2} (e^1_{a \mu b} - e^1_{b \mu a}),
\eqno (5.6a) $$

which is the usual expression for the covariant momenta in
gauge gravitational fields \cite{wess}.
One can find  the two constraints

$$ p_0^\mu p_{0 \mu} = 0 = p_{0 \mu} f^{1 \mu} { }_a \tilde{a}^a .
\eqno (5.7)$$

To see  how  Yang-Mills fields enter into the theory,
the Dirac equation (5.7) has to be rewritten in terms of fields
which determine
gravitation in the four dimensional subspace and of those fields
which determine  gravitation in higher dimensions, assuming
that the coordinates of ordinary space with indices higher than
four  stay compacted to  unmeasurably small dimensions.
Since  Grassmann space manifests itselfs through  average
values of observables only, compactification of a part of
Grassmann space has no meaning.  However, since
parameters of  Lorentz transformations in a freely fallying
coordinate system for both spaces have
to be the same, no transformations to the fifth or
higher coordinates may occur
at measurable energies. Therefore, the four dimensional subspace
of  Grassmann space with the generators defining the Lorentz
group $ SO(1,3)$ is (almost) decomposed from the rest of the
Grassmann space with the generators forming the (compact) group
$ SO(d-4) $, because of the decomposition of  ordinary
space. This is valide on the classical level only.

We shall assume the case in which  only some components
of fields differ from zero:

$$ \left( \begin{array}{cc|cc}
   e^{im}{ }_{\alpha}& & & 0\\
   & & &\\ \hline
   & & &\\
   0 & & & e^{ih}{ }_{\sigma}
\end{array} \right), \;\;\;\;
\alpha,m \in (0,3),\: \sigma,h \in (5,d),\: i \in (1,2), \eqno
(5.8)$$

while vielbeins $ e^{im}{ }_{\alpha}, e^{ik}{ }_{\sigma}$ depend
on $ \theta^a and x^{\alpha}, \alpha \in{0,3}$ only (since the
dependence on other coordinates is not measurable ).
Accordingly we have ( see
Eq.(5.4)) only $ \omega _{ab \alpha} \neq 0,$. We recognize, as
in the freely fallying coordinate system,
that Grassmann coordinates with indices
from $ 0 $ to $ 5 $ determine spins of fields, while Grassmann
coordinates with indices higher than $5$ determine charges of fields.
We shall take  expectation values of $
p^h = 0 $ ,  $ a \ge 6 $  while we take $ m = - p_5 ( e^{15}
{ }_5 ) { }^{-1} $. We find

$$ \tilde{ \gamma}^a f^{1 \mu} { }_a p_{0 \mu} = \tilde{ \gamma}
^m f^{1 \alpha} { }_m ( p_{ \alpha} - \frac{1}{2} \tilde{S} ^{mn}
 \omega_{mn \alpha} +\hbox{\it }A_{\alpha} ) + m, \;\;
\hbox{ \rm where} \hbox{\it A}_{\alpha} = \sum_{A,i}
\tilde{ \tau}^{Ai} \hbox{\it
A}^{Ai}_{\alpha},
\eqno(5.9) $$

with
$\sum_{A,i} \tilde{ \tau}^{Ai} \hbox{\it A}^{Ai}_{ \alpha} =
\frac{1}{2}  \tilde{S}^{hk}  \omega_{hk \alpha} ,
 \;\;h,k = 6,7,8,..d. $

According to
Eqs.(3.1) the Lie algebra of the Lorentz group $ SO ( d-5
) $ contains the appropriate subalgebras
 for the desired charges  if
$\tilde{ \tau}^{Ai}$ can be expressed as linear
superpositions  of operators $ \tilde S ^{ab}, $
and $ f^{Aijk}$ are structure constants of the $n$ subgroups
$A$ , each with $ n_A$ operators.

As we already stated in Section 2, this can certainly  be done for
$d = 15 $,
since $ SO(1,14) $ has
the subalgebras $ SO(1,4)\times SO(10) $, while $ SO(10) $ has
the subalgebras $ SU(3)\times SU(2)\times U(1) $.

We further find, if the subalgebras of operators $
\tilde{\tau}^{Ai} $
are isomorphic to the algebra of $ SO(d-5) $, that
$ \hbox{\it A}^{Ai}{ }_\alpha = \frac{1}{2} \omega_{hk \alpha}
\hbox{\it d}^{Aihk}. $
Subalgebras of $ SO(10) $ and $ SU(3)\times SU(2)\times U(1) $
are not isomorphic. In this case there follow $ \frac
{10.(10-1)}{2} $ equations, which determine the same number of
fields $ \omega_{hk \alpha} $, for each $ \alpha = 0,1,2,3 $, in
terms of $ 8 + 3 + 1 $ fields $ \hbox{\it A}^ {Ai}{ }_{\alpha} $.
When these equations are fulfilled, the symmetry $ SO(10) $ is
broken to $ SU(3)\times SU(2)\times U(1) $.

In eq.(5.9) the fields $ \omega _{hk \alpha} $ determine all
the Yang-Mills fields, including electromagnetic ones. The
proposed unification differs from the Kaluza-Klein types of
unification, since Yang-Mills fields are not determined by
nondiagonal terms of vielbeins $ e^{ih}{ }_{\alpha} $. Instead
they are determined with spin connections.
In the proposed theory there is no difficulties either with the
Planck mass of the electron ( since the electron's charge is not
determined by the momentum $ p^{\sigma}, \sigma=5 $ but with the
generators of Lorentz transformations in Grassmann space ) or
with  transformation properties of gauge fields.

Torsion and  curvature follow from the Poisson brackets
$ \{ p_{0a},  p_{0b} \}_p $, with $ p_{0a} = f^{1\mu} { }_{a} ( p_{\mu} -
\frac{1}{2} \tilde{S}^{cd} \omega_{cd\mu}) $.

We find
$ \{ p_{0 a} , p_{0 b} \}_p = -\frac{1}{2} S^{c d} R_{ c d a b}
+ p_{0 c} T^c{ }_{a b} ,$
$ R_{c d a b} = f^{1 \mu} {}_{[a} f^{1 \nu} {}_{b]}
( \omega_{cd\nu,\mu^{x} } + \omega_{c} {}^e{}_{ \mu} \omega_{e d \nu}
+ \overrightarrow{\omega}_{c d \mu , f^{\theta}} \theta^e \omega_e
{ }^f{ }_{ \nu}), $
$ T^c { }_{a b}  = e^{1c} { }_{ \mu} ( f^{1 \nu} {}_{[b} f^{1 \mu}
{}_{ a]}  {}_{,
\nu} + \omega_{e \nu} { }^d \theta^e f^{1 \nu} {}_{[b}
\overrightarrow{f^{1 \mu}} {}_{  a]} {}_{, d^\theta} ),$
 with $ \; A _{[a} B _{b]} $  $ = A_a B_b - A_b B_a. $

For $e^{im}{ }_{\alpha} = \delta ^m{ }_{\alpha}$ one easily sees
that Eq.(5.9) manifests
the Dirac equation for a paricle with Yang-Mills charges in
external fields.

While $ \tilde S^{ab} $ determine the fundamental
representations of the Lorentz group and therefore define $\tilde
S^{mn},\; m,n \in {0,3} $  spins of fermionic
fields and $\tilde S^{h,k},\, h,k \in {6,d} $  their
Yang-Mills charges, determine accordingly $ S^{a,b} $ adjoint
representations of the Lorentz group and therefore define $ S^{m,n},
m,n \in {0,3} $  spins of bosonic fields and $ S^{h,k}, h,k \in
{6,d} $ their charges.

\vskip 1cm

\noindent
{\bf CONCLUDING REMARKS}

\vskip 0.5cm

In this talk the theory in which space has d ordinary and d
Grassmann coordinates was presented. Two kinds of generators of
Lorentz tranformations in Grassmann space can be defined.
The generators of spinorial character define the fundamental
representations of the Lorentz group, the generators of the
vectorial character define the adjoint representations of the
Lorentz group. Both kinds of generators are the linear
differential operators in Grassmann space. The Lorentz group $
SO(1,d-1)$ contains for $d=15$ as subgroups $ SO(1,4), SU(3),
SU(2) $ and $ U(1) $. While $ SO(1,4)$ defines spins of fermionic
and bosonic fields, define $ SU(3), SU(2)$ and $ U(1)$ charges
of both fields. Charges of fermionic fields belong to the
fundamental representations, while charges of bosonic fields
belong to the adjoint representations.

When looking for the representations of the operators $ \tilde
{S}^{mn},\; m,n\in{0,3} $ as polynomials of $ \theta^a,\; a \in{0,5}$
and operators $\tilde{\tau}^{Ai} $ as polynomials of $ \theta^h,
\;a \in{6,15}$, we find  representations of the group
$SO(1,14)$ as  outer products of the representations of
subgroups. The Grassmann odd polynomials, which are the Dirac
four spinors, are triplets or singlets with respect to the
colour charge, doublets or singlets with respect to the weak
charge and may have hypercharge equal to $\pm \frac{1}{6}, \pm
\frac {1}{3}, \pm \frac {2}{3}, \pm \frac {1}{2}, \pm 1, 0 $.

When looking for the representations of $ SO(1,14)$, as
Grassmann even polynomials of $ \theta^a $, in terms of
the subgroups $ SO(1,4), SU(3), SU(2), U(1) $, we find scalars
and vectors, which are singlets, octets and multiplets with
fourteen vectors with respect to the colour charge, triplets and
singlets with respect to the weak charge and may have the
hypercharge equal to $ 0 $ or to $\pm 1$. These representations
are presented in Ref. \cite{man2}.

We presented the Lagrange function for a particle living on a
supergeodesics, with the momentum in the Grassmann space
proportional to the Grassmann coordinate. In the quantization
procedure the Dirac equation follows, with $\gamma^a $
operators, which have the odd Grassmann character and are
differential operators in  Grassmann space with coordinates
$\theta^a,\;a \in{0,5} $. When transforming the Lagrange
function from the freely fallyng to the external coordinate
system, vielbeins and spin connections describe not only the
gravitational field but also the Yang-Mills fields. Since the
generators of the Lorentz transformations with indices higher
then five determine charges of particles and spin connections
again with indices higher then five describe the Yang-Mills
fields (rather then vielbeins with one index smaller then five
and another greater then three as in the Kaluza-Klein theories),
the problems of the Kaluza-Klein theories ( like a Planck mass of
charged particles ) do not occur.

\vskip 0.5cm

\noindent
{\bf ACKNOWLEDGEMENT. } This work was supported by Ministry of
Science and Technology of Slovenia.


\end{document}